\documentclass[12pt,preprint]{aastex} 
\usepackage{graphicx,color}
\usepackage{amsmath,amssymb}

\newcommand\unit[1]{\ \mathrm{#1}}

\def\AU{\unit{AU}}

\def\My{\unit{My}}
\def\Gy{\unit{Gy}}

\def\yr{\unit{y}}

\def\Mearth{\unit{M}_\oplus} 
\def\erf{\mathrm{erf}}

\begin{document} 
\title{Secular resonance sweeping of the main asteroid belt during planet migration}
\author{David A. Minton}
\affil{Southwest Research Institute and NASA Lunar Science Institute}
\affil{1050 Walnut St., Suite. 300, Boulder, CO 80302}
\email{daminton@boulder.swri.edu}
\and 
\author{Renu Malhotra}
\affil{Department of Planetary Sciences, University of Arizona}
\affil{1629 East University Boulevard, Tucson, AZ 85721}
\email{renu@lpl.arizona.edu}

\slugcomment{32 pages, 7 figures. Accepted for publication in ApJ on March 1, 2011}

\begin{abstract}
We calculate the eccentricity excitation of asteroids produced by the sweeping $\nu_6$ secular resonance during the epoch of planetesimal-driven giant planet migration in the early history of the solar system.  
We derive analytical expressions for the magnitude of the eccentricity change and its dependence on the sweep rate and on planetary parameters; the $\nu_6$ sweeping leads to either an increase or a decrease of eccentricity depending on an asteroid's initial orbit.  
Based on the slowest rate of $\nu_6$ sweeping that allows a remnant asteroid belt to survive, we derive a lower limit on Saturn's migration speed of $\sim0.15\AU\My^{-1}$ during the era that the $\nu_6$ resonance swept through the inner asteroid belt (semimajor axis range $2.1$--$2.8\AU$). This rate limit is for Saturn's current eccentricity, and scales with the square of Saturn's eccentricity; the limit on Saturn's migration rate could be lower if Saturn's eccentricity were lower during its migration.
Applied to an ensemble of fictitious asteroids, our calculations show that a prior single-peaked distribution of asteroid eccentricities would be transformed into a double-peaked distribution due to the sweeping of the $\nu_6$.  
Examination of the orbital data of main belt asteroids reveals that the proper eccentricities of the known bright ($H \leq10.8$) asteroids may be consistent with a double-peaked distribution. 
If so, our theoretical analysis then yields two possible solutions for the migration rate of Saturn and for the dynamical states of the pre-migration asteroid belt: a dynamically cold state (single-peaked eccentricity distribution with mean of $\sim0.05$) linked with Saturn's migration speed  $\sim 4\AU\My^{-1}$, or a dynamically hot state (single-peaked eccentricity distribution with mean of $\sim0.3$) linked with Saturn's migration speed $\sim 0.8\AU\My^{-1}$.

\end{abstract} 


\section{Introduction}
\label{sec:sweeprate-intro}
The dynamical structure of the Kuiper Belt suggests that the outer solar system experienced a phase of planetesimal-driven migration in its early history~\citep{Fernandez:1984p61,Malhotra:1993p244,Malhotra:1995p79,Hahn:1999p122,Levison:2008p690}.  
Pluto and other Kuiper belt objects that are trapped in mean motion resonances (MMRs) with Neptune are explained by the outward migration of Neptune due to interactions with a more massive primordial planetesimal disk in the outer regions of the solar system~\citep{Malhotra:1993p244,Malhotra:1995p79}.
In addition, the so-called the scattered disk of the Kuiper belt can also be explained by the outward migration of Neptune~\citep{Hahn:2005p4122}, or by the effects of a high eccentricity phase of ice giant planet evolution during the outward migration of Neptune~\citep{Levison:2008p690}.
The basic premise of planetesimal-driven migration is that the giant planets formed in a more compact configuration than we find them today, and that they were surrounded by a massive ($\sim50\Mearth$) disk of unaccreted icy planetesimals that was the progenitor of the currently observed Kuiper belt~\citep{Hahn:1999p122}.
When planetesimals are preferentially scattered either inward (toward the Sun) or outward (away from the Sun), net orbital angular momentum is transferred between the disk and the large body, causing a drift in the large body's semimajor axis~\citep{Fernandez:1984p61,Kirsh:2009p2746}.
In many simulations of giant planet migration, icy planetesimals are preferentially scattered inward by each of the three outer giant planets (Saturn, Uranus, and Neptune) causing these planets to migrate outward. 
Due to Jupiter's large mass, planetesimals that encounter Jupiter are preferentially ejected out of the solar system, leading to a net loss of mass from the solar system and an inward migration of Jupiter.

Planetesimal-driven giant planet migration has been suggested as a cause of the Late Heavy Bombardment (LHB)~\citep{Gomes:2005p51,Strom:2005p80}, however the link between these two events has yet to be definitively established~\citep{Chapman:2007p97,Cuk:2010p4126,Malhotra:2010}.
Such migration would have enhanced the impact flux of both asteroids and comets onto the terrestrial planets in two ways. 
First, many of the icy planetesimals scattered by the giant planets would have crossed the orbits of the terrestrial planets.  
Second, as the giant planets migrated, locations of mean motion and secular resonances would have swept across the asteroid belt, raising the eccentricities of asteroids to planet-crossing values. 

Recently, \cite{Minton:2009p280} showed that the patterns of depletion observed in the asteroid belt are consistent with the effects of sweeping of resonances during the migration of the giant planets.
The Jupiter-facing sides of some of the Kirkwood gaps (regions of the asteroid belt that are nearly empty due to strong jovian mean motion resonances) are depleted relative to the Sun-facing sides, as would be expected due to the inward migration of Jupiter and the associated inward sweeping of the jovian mean motion resonances.
The region within the inner asteroid belt between semimajor axis range $2.1$--$2.5\AU$ also has excess depletion relative to a model asteroid belt that was uniformly populated and then subsequently sculpted by the gravitational perturbations of the planets over $4\Gy$, as would be expected due to the outward migration of Saturn and the associated inward sweeping of a strong secular resonance, the so-called $\nu_6$ resonance, as explained below.
In our 2009 study, we concluded that the semimajor axis distribution of asteroids in the main belt is consistent with the inward migration of Jupiter and outward migration of Saturn by amounts proposed in previous studies based on the Kuiper belt resonance structure\citep[e.g.,~][]{Malhotra:1995p79}.
% In limited numerical experiments reported in that study, we found that the asteroids' semimajor axis distribution is consistent with an e-folding migration timescale $\tau = 0.5$ Myr.
However, in that study the migration timescale was not strongly constrained, because only the relative depletion of asteroids in nearby semimajor axis bins could be determined, not their overall level of depletion. 
In the present paper, we explore in more detail the effect that planet migration would have had on the asteroid belt due to asteroid eccentricity excitation by the sweeping of the $\nu_6$ secular resonance.  
From the observed eccentricity distribution of main belt asteroids, we find that it is possible to derive constraints on the secular resonance sweeping timescale, and hence on the migration timescale.

Secular resonances play an important role in the evolution of the main asteroid belt.  
The inner edge of the belt nearly coincides with the $\nu_6$ secular resonance which is defined by $g\approx g_6$, where $g$ is the rate of precession of the longitude of pericenter, $\varpi$, of an asteroid and $g_6$ is the sixth eigenfrequency of the solar system planets (approximately the rate of precession of Saturn's longitude of pericenter).
The $\nu_6$ resonance, is important for the delivery of Near Earth Asteroids (NEAs) to the inner solar system~\citep{Scholl:1991p680}.
\cite{Williams:1981p532} showed that the location of the $\nu_6$ resonance actually forms surfaces in $a-e-\sin i$ space, and \cite{Milani:1990p531} showed that those surfaces approximately define the ``inner edge" of the main asteroid belt.
However, as mentioned above, \cite{Minton:2009p280} found that, with the planets in their present configuration, planetary perturbations over the age of the solar system cannot fully account for the detailed orbital distribution of the asteroids in the inner asteroid belt.
The pattern of excess depletion in inner asteroid belt noted by \cite{Minton:2009p280} is consistent with the effect of the inward sweeping of the $\nu_6$ secular resonance.
In general, the direction of motion of the $\nu_6$ is anticorrelated with that of Saturn, so an inward sweeping of the $\nu_6$ would be produced by an outwardly migrating Saturn.

Sweeping, or scanning, secular resonances have been analyzed in a number of previous works.  
Sweeping secular resonances due to the changing quadrupole moment of the Sun during solar spin-down have been explored as a possible mechanism for explaining the eccentricity and inclination of Mercury~\citep{Ward:1976p129}.
Secular resonance sweeping due to the effects of the dissipating solar nebula just after planet formation has also been investigated as a possible mechanism for exciting the orbital eccentricities of Mars and of the asteroid belt~\citep{Heppenheimer:1980p32,Ward:1981p69}.
The dissipating massive gaseous solar nebula would have altered the secular frequencies of the solar system planets in a time-dependent way, causing locations of secular resonances to possibly sweep across the inner solar system, thereby exciting asteroids into the eccentric and inclined orbits that are observed today.  
This mechanism was revisited by \cite{Nagasawa:2000p2239}, who incorporated a more sophisticated treatment of the nebular dispersal.  
However, \cite{OBrien:2007p95} have argued that the excitation (and clearing) of the primordial asteroid belt was unlikely due to secular resonance sweeping due to the dispersion of the solar nebula.

The special case of asteroids on initially circular orbits being swept by the $\nu_6$ and $\nu_{16}$ resonances has been investigated by~\cite{Gomes:1997p1016}.  
In this paper, we consider the more general case of non-zero initial eccentricities; our analysis yields qualitatively new results and provides new insights into the dynamical history of the asteroid belt.  
This extends the work of \cite{Ward:1976p129} and \cite{Gomes:1997p1016} in developing analytical treatments of the effects of sweeping secular resonances on asteroid orbits.   
In doing so, we have developed an explicit relationship between the migration rate of the giant planets, the initial eccentricity of the asteroid and its initial longitude of perihelion, and the final eccentricity of the asteroid after the passage of the resonance.  
We show that for initially non-zero asteroid eccentricity, the sweeping of the $\nu_6$ resonances can either increase or decrease asteroid eccentricities. 
Examining the orbits of observed main belt asteroids we find evidence for a double-peaked eccentricity distribution; this supports the case for a history of $\nu_6$ sweeping.   
Quantitative comparison of our analytical theory with the semimajor axis and eccentricity distribution of asteroids yields new constraints on the timescale of planet migration.

We note that although our analysis is carried out in the specific context of the sweeping $\nu_6$ resonance during the phase of planetesimal-driven migration of Jupiter and Saturn, the techniques developed here may be extended to other similar problems, for example, the sweeping of the inclination-node $\nu_{16}$ resonance in the main asteroid belt, the $\nu_8$ secular resonance in the Kuiper belt, and farther afield, the sweeping of secular resonances in circumstellar or other astrophysical disks.

\section{Analytical theory of a sweeping secular resonance}
\label{sec:sweeprate-toy_model}
We adopt a simplified model in which a test particle (asteroid) is perturbed only by a single resonance, the $\nu_6$ resonance. 
We use a system of units where the mass is in solar masses, the semimajor axis is in units of AU, and the unit of time is $(2\pi)^{-1}$y.
With this choice, the gravitational constant, $G$, is unity.
An asteroid's secular perturbations close to a secular resonance can be described by the following Hamiltonian function~\citep{Malhotra:1998p124}:
\begin{equation}
    H_{sec}=-g_0J+\varepsilon\sqrt{2J}\cos(w_p-\varpi)\label{e:resHamiltonian},
\end{equation}
where $w_p=g_pt+\beta_p$ describes the phase of the p-th eigenmode of the linearized eccentricity-pericenter secular theory for the Solar system planets~\citep{Murray:1999SSD}, $g_p$ is the associated eigenfrequency, $\varpi$ is the asteroid's longitude of perihelion, 
$J=\sqrt{a}\left(1-\sqrt{1-e^2}\right)$ is the canonical generalized momentum which is related to the asteroid's orbital semimajor axis $a$ and eccentricity $e$; $-\varpi$ and $J$ are the canonically conjugate pair of variables in this 1-degree-of-freedom Hamiltonian system.   
The coefficients $g_0$ and $\varepsilon$ are given by:
\begin{align}
        g_0&=\frac{1}{4a^{3/2}}\sum_j\alpha_j^2b^{(1)}_{3/2}(\alpha_j)m_j\label{e:g0def},\\
    \varepsilon&=\frac{1}{4a^{5/4}}\sum_j\alpha_j^2b^{(2)}_{3/2}(\alpha_j)m_jE_j^{(p)},
\label{e:epsdef}
\end{align}
where the subscript $j$ refers to a planet, $E_j^{(p)}$ is the amplitude of the $g_p$ mode in the $j^{th}$ planet's orbit, $\alpha_j=\min\{a/a_j, a_j/a\}$, $m_j$ is the ratio of the mass of planet $j$ to the Sun, and $b_{3/2}^{(1)}(\alpha_j)$ and $b_{3/2}^{(2)}(\alpha_j)$ are Laplace coefficients; the sum is over all major planets.  
The summations in equations~(\ref{e:g0def})--(\ref{e:epsdef}) are over the 8 major planets, for the greatest accuracy; however, we will adopt the simpler two-planet model of the Sun-Jupiter-Saturn in \S\ref{s:sweeping}, in which case we sum over only the indices referring to Jupiter and Saturn; then $g_p$ is an eigenfrequency of the secular equations of the two planet system.

With fixed values of the planetary masses and semimajor axes, $g_0$, $g_{\rm p}$ and $\varepsilon$ are constant parameters in the Hamiltonian given in equation~(\ref{e:resHamiltonian}).  
However, during the epoch of giant planet migration, the planets' semimajor axes change secularly with time, so that  $g_0$, $g_{\rm p}$ and $\varepsilon$ become time-dependent parameters.  
In the analysis below, we neglect the time-dependence of $g_0$ and $\varepsilon$, and adopt a simple prescription for the time-dependence of $g_{\rm p}$ (see equation~\ref{e:goft} below).  
This approximation is physically motivated: the fractional variation of $g_0$ and $\varepsilon$ for an individual asteroid is small compared to the effects of the ``small divisor" $g_0-g_{\rm p}$ during  the $\nu_6$ resonance sweeping event.

It is useful to make a canonical transformation to new variables $(\phi, P)$ defined by the following generating function,
\begin{equation}
    {\cal F}(-\varpi,P,t) = (w_p(t)-\varpi)P
\end{equation}
Thus, $\phi = \partial{\cal F}/\partial P = (w_p(t)-\varpi)$ and $J=-\partial{\cal F}/\partial\varpi=P$.  
The new Hamiltonian function is $\tilde H_{sec}=H_{sec}+\partial{\cal F}/\partial t$,
\begin{equation}
    \tilde H_{sec}=(\dot w_p(t)-g_0)J+\varepsilon\sqrt{2J}\cos\phi, \label{e:resHamiltonian2}
\end{equation}
where we have retained $J$ to denote the canonical momentum, since $P=J$.  
It is useful to make a second canonical transformation to canonical eccentric variables,
\begin{equation}
   x = \sqrt{2J}\cos\phi, \qquad y=-\sqrt{2J}\sin\phi, 
\end{equation}
where $x$ is the canonical coordinate and $y$ is the canonically conjugate momentum.  
The Hamiltonian expressed in these variables is
\begin{equation}
   \tilde H_{sec}=(\dot w_p(t)-g_0)\frac{x^2+y^2}{2}+\varepsilon x.\label{e:resHamiltonian3}
\end{equation}

As discussed above, during planetary migration, the secular frequency $g_p$ is a slowly varying function of time.  
We approximate its 
rate of change, $\dot g_p=2\lambda$,  as a constant, so that
\begin{equation}
	\dot w_p(t) =g_{p,0}+2\lambda t.
	\label{e:goft}
\end{equation} 
We define $t=0$ as the epoch of exact resonance crossing, so that $g_{p,0}=g_0$~\citep[cf.~][]{Ward:1976p129}.
Then, $\dot w_p(t)-g_0=2\lambda t$, and the equations of motion from the Hamiltonian of equation~(\ref{e:resHamiltonian3}) can be written as:
\begin{align}
	\dot{x}&=2\lambda t y,\label{e:dotxsweep}\\
	\dot{y}&=-2\lambda t x-\varepsilon.\label{e:dotysweep}
\end{align}
These equations of motion form a system of linear, nonhomogenous differential equations, whose solution is a linear combination of a homogeneous and a particular solution. 
The homogeneous solution can be found by inspection, giving:
\begin{align}
	x_h(t)&=c_1\cos\lambda t^2+c_2\sin\lambda t^2,\label{e:xhomog}\\
	y_h(t)&=-c_1\sin\lambda t^2+c_2\cos\lambda t^2\label{e:yhomog},
\end{align}
where $c_1$ and $c_2$ are constant coefficients.  
We use the method of variation of parameters to find the particular solution.
Accordingly, we replace the constants $c_1$ and $c_2$ in the homogeneous solution with functions $A(t)$ and $B(t)$, to seek the particular solution of the form
\begin{align}
	x_p(t)&=A(t)\cos\lambda t^2+B(t)\sin\lambda t^2\label{e:xp}\\
	y_p(t)&=-A(t)\sin\lambda t^2+B(t)\cos\lambda t^2.\label{e:yp}
\end{align}
Substituting this into the equations of motion we now have:
\begin{align}
	&\dot{A}\cos\lambda t^2+\dot{B}\sin\lambda t^2=0,\label{e:xvarparam}\\
	-&\dot{A}\sin\lambda t^2+\dot{B}\cos\lambda t^2=-\varepsilon;\label{e:yvarparam}
\end{align}
therefore
\begin{align}
	\dot{A}&=\varepsilon\sin\lambda t^2,\label{e:dotA}\\
	\dot{B}&=-\varepsilon\cos\lambda t^2.\label{e:dotB}
\end{align}
Equations~(\ref{e:dotA}) and (\ref{e:dotB}) do not have a simple closed-form solution, but their solution can be expressed in terms of Fresnel integrals~\citep{Zwillinger:1996CRC}.  
The Fresnel integrals are defined as follows:
\begin{align}
	S(t)&=\int_0^t\sin t'^2 dt',\label{e:Sfresnel}\\
	C(t)&=\int_0^t\cos t'^2 dt'.\label{e:Cfresnel}
\end{align}
and have the following properties:
\begin{align}
	S(-t)&=-S(t)\label{e:negSfresnellimit},\\
	C(-t)&=-C(t)\label{e:negCfresnellimit},\\
	S(\infty)&=C(\infty)=\sqrt{\frac{\pi}{8}}.\label{e:fresnellimits}
\end{align}
Therefore
\begin{align}
	A(t)&= \frac{\varepsilon}{\sqrt{|\lambda|}}S\left(t\sqrt{|\lambda|}\right),\label{e:Afresnel}\\
	B(t)&=-\frac{\varepsilon}{\sqrt{|\lambda|}}C\left(t\sqrt{|\lambda|}\right).\label{e:Bfresnel}
\end{align}

We denote initial conditions with a subscript $i$, and write the solution to equations~(\ref{e:dotxsweep}) and (\ref{e:dotysweep}) as
\begin{align}
	x(t)&=x_i\cos\left[\lambda\left(t^2-t_i^2\right)\right]+y_i\sin\left[\lambda\left(t^2-t_i^2\right)\right]\notag\\
		 &\qquad+\frac{\varepsilon}{\sqrt{|\lambda|}}\left[\left(S-S_i\right)\cos\lambda t^2-\left(C-C_i\right)\sin\lambda t^2\right]\label{e:xsoln},\\
	y(t)&=-x_i\sin\left[\lambda\left(t^2-t_i^2\right)\right]+y_i\cos\left[\lambda\left(t^2-t_i^2\right)\right]\notag\\
		 &\qquad-\frac{\varepsilon}{\sqrt{|\lambda|}}\left[\left(C-C_i\right)\cos\lambda t^2+\left(S-S_i\right)\sin\lambda t^2\right]\label{e:ysoln}.
\end{align}
Because the asteroid is swept over by the secular resonance at time $t=0$, we can calculate the changes in $x,y$ by 
letting $t_i=-t_f$ and evaluating the coefficients $C_i,C_f, S_i, S_f$ far from resonance passage, i.e., for $t_f\sqrt{|\lambda|}\gg1$, by use of equation~(\ref{e:fresnellimits}).  
Thus we find
\begin{align}
	x_f&=x_i+\varepsilon\sqrt{\frac{\pi}{2|\lambda|}}\left[\cos\lambda t_i^2-\sin\lambda t_i^2\right],\label{e:xf}\\
	y_f&=y_i- \varepsilon\sqrt{\frac{\pi}{2|\lambda|}}\left[\cos\lambda t_i^2+\sin\lambda t_i^2\right].\label{e:yf}
\end{align}

The final value of $J$ long after resonance passage is therefore given by 
\begin{align}
	J_f&=\frac{1}{2}\left(x_f^2+y_f^2\right)\notag\\
	     &=\frac{1}{2}\left(x_i^2+y_i^2\right)+\frac{\pi\varepsilon^2}{2|\lambda|} + \varepsilon\sqrt\frac{\pi}{2|\lambda|}[x_i(\cos\lambda t_i^2-\sin\lambda t_i^2)\notag\\
	&\qquad-y_i(\cos\lambda t_i^2+\sin\lambda t_i^2)]\notag\\
              &=J_i+\frac{\pi\varepsilon^2}{2|\lambda|}+\varepsilon\sqrt{\frac{2\pi J_i}{|\lambda|}}\cos(\phi_i-\lambda t_i^2-\frac{\pi}{4}).\label{e:Jftf}
\end{align}
With a judicious choice of the initial time, $t_i$, and without loss of generality, the cosine in the last term becomes $\cos\varpi_i$, and therefore 
\begin{equation}
	J_f=J_i+\frac{\pi\varepsilon^2}{2|\lambda|}+\varepsilon\sqrt{\frac{2\pi J_i}{|\lambda|}}\cos\varpi_i\label{e:Jf}.
\end{equation}

The asteroid's semimajor axis $a$ is unchanged by the secular perturbations; thus, the changes in $J$ reflect changes in the asteroid's eccentricity $e$.  
For asteroids with non-zero initial eccentricity, the phase dependence in equation~(\ref{e:Jf}) means that secular resonance sweeping can potentially both excite and damp orbital eccentricities.  
We also note that the magnitude of eccentricity change is inversely related to the speed of planet migration.  
In linear secular theory (equation~(\ref{e:resHamiltonian})), eccentricity and inclination are decoupled, and therefore the effect of the sweeping $\nu_6$ does not depend on the inclination.
However, as \cite{Williams:1981p532} showed, the location of the $\nu_6$ does depend on inclination, but the dependence is weak for typical inclinations of main belt objects.  
Nevertheless, there are populations of main belt asteroids at high inclination (such as the Hungaria and Phocaea families), and an analysis of secular resonance sweeping that incorporates coupling between eccentricity and inclination would be valuable for understanding the effects of planet migration on these populations; we leave this to a future investigation.

For small $e$, we can use the approximation $J\simeq{\frac{1}{2}}\sqrt{a}e^2$. 
Considering all possible values of $\cos\varpi_i\in\{-1,+1\}$, an asteroid with initial eccentricity $e_i$ that is swept by the $\nu_6$ resonance will have a final eccentricity in the range $e_{min}$ to $e_{max}$, where
\begin{equation}
	e_{min,max}\simeq\left|e_i\pm\delta_e\right|,
	\label{e:finalebounds}
\end{equation}
and 
\begin{equation}
	\delta_e\equiv\left|\varepsilon\sqrt{\frac{\pi}{|\lambda|\sqrt{a}}}\right|\label{e:deltaedef}.
\end{equation}

Equations~(\ref{e:Jf})--(\ref{e:deltaedef}) have the following implications: 
\begin{enumerate}
\item Initially circular orbits become eccentric, with a final eccentricity $\delta_e$.  
\item An ensemble of orbits with the same $a$ and initial non-zero $e$ but uniform random orientations of pericenter are transformed into an ensemble that has eccentricities in the range $e_{min}$ to $e_{max}$; this range is not uniformly distributed because of the $\cos\varpi_i$ dependence in equation~(\ref{e:Jf}), rather the distribution peaks at the extreme values (see Fig.~\ref{f:tpideal} below).  
\item An ensemble of asteroids having an initial distribution of eccentricities which is a single-peaked Gaussian (and random orientations of pericenter) would be transformed into one with a double-peaked eccentricity distribution.  
\end{enumerate}

\section{Application to the Main Asteroid Belt}
\label{s:sweeping}

In light of the above calculations, it is possible to conclude that if the asteroid belt were initially dynamically cold, that is asteroids were on nearly circular orbits prior to secular resonance sweeping, then the asteroids would be nearly uniformly excited to a narrow range of final eccentricities, the value of which would be determined by the rate of resonance sweeping.  
Because asteroids having eccentricities above planet-crossing values would be unlikely to survive to the present day, it follows that an initially cold asteroid belt which is uniformly excited by the $\nu_6$ sweeping will either lose all its asteroids or none.   
On the other hand, an initially excited asteroid belt, that is a belt with asteroids that had non-zero eccentricities prior to the $\nu_6$ sweeping, would have asteroids' final eccentricities bounded by equation~(\ref{e:finalebounds}), allowing for partial depletion and also broadening of its eccentricity distribution.   In this section, we apply our theoretical analysis to the problem of the $\nu_6$ resonance sweeping through the asteroid belt, and compare the theoretical predictions with the observed eccentricity distribution of asteroids.

\subsection{Parameters} In order to apply the theory, we must find the location of the $\nu_6$ resonance as a function of the semimajor axes of the giant planets orbits, and also obtain values for the parameter $\varepsilon$ (equation~\ref{e:epsdef}), for asteroids with semimajor axis values in the main asteroid belt. 
The location of the $\nu_6$ resonance is defined as the semimajor axis, $a_{\nu_6}$, where the rate, $g_0$ (equation~\ref{e:g0def}), 
of pericenter precession of a massless particle (or asteroid) is equal to the $g_6$ eigenfrequency of the solar system.  
In the current solar system, the $g_6$ frequency is associated with the secular mode with the most power in Saturn's eccentricity--pericenter variations.  
During the epoch of planetesimal-driven planet migration, Jupiter migrated by only a small amount but Saturn likely migrated significantly more~\citep{Fernandez:1984p61,Malhotra:1995p79,Tsiganis:2005p39}, so we expect that the variation in location of the $\nu_6$ secular resonance is most sensitive to Saturn's semimajor axis. 
We therefore adopt a simple model of planet migration in which Jupiter is fixed at $5.2\AU$ and only Saturn migrates.  
We neglect the effects of the ice giants Uranus and Neptune, as well as secular effects due to the previously more massive Kuiper belt and asteroid belt.  
With these simplifications, the $g_6$ frequency varies with time as Saturn migrates, so we parametrize $g_6$ as a function of Saturn's semimajor axis.  
In contrast with the variation of $g_6$, there is little variation of the asteroid's apsidal precession rate, $g_0$, as Saturn migrates.  Thus, finding $a_{\nu_6}$ is reduced to calculating the dependence of $g_6$ on Saturn's semimajor axis.

To calculate $g_6$ as a function of Saturn's semimajor axis, we proceed as follows.
For fixed planetary semimajor axes, the Laplace-Lagrange secular theory provides the secular frequencies and orbital element variations of the planets.
This is a linear perturbation theory, in which the disturbing function is truncated to secular terms of second order in eccentricity and first order in mass~\citep{Murray:1999SSD}.  
In the planar two-planet case, the secular perturbations of planet $j$, where $j=5$ is Jupiter and $j=6$ is Saturn, are described by the following disturbing function:
\begin{equation}
	R_{j}=\frac{n_j}{a_j^2}\left[\frac{1}{2}A_{jj}e_j^2+A_{jk}e_5e_6\cos(\varpi_5-\varpi_6)\right],
	\label{e:secHam}
\end{equation}
where $n$ is the mean motion, and $\mathbf{A}$ is a matrix with elements
\begin{eqnarray}
	A_{jj}&=+n_j\frac{1}{4}\frac{m_k}{M_\odot+m_j}\alpha_{jk}\bar{\alpha}_{jk}b_{3/2}^{(1)}(\alpha_{jk}),\label{e:Ajjelement}\\
	A_{jk}&=-n_j\frac{1}{4}\frac{m_k}{M_\odot+m_j}\alpha_{jk}\bar{\alpha}_{jk}b_{3/2}^{(2)}(\alpha_{jk}),\label{e:Ajkelement}
\end{eqnarray}
for $j=5,6$, $k=6,5$, and $j\ne k$; $\alpha_{jk}=\min\{a_j/a_k,a_k/a_j\}$, and 
\begin{equation}
	\bar{\alpha}_{jk}=\left\{
     \begin{array}{lr}
       1 & : a_j>a_k \\
       a_j/a_k & : a_j<a_k. \\
     \end{array}
   \right.
	\label{e:baralpha}
\end{equation}
The secular motion of the planets is then described by a set of linear differential equations for the eccentricity vectors, $e_j(\sin\varpi_j,\cos\varpi_j)\equiv(h_j,k_j)$, \begin{equation}
	\dot{h}_j=+\sum_{p=5}^6A_{pj}k_j,\;\;\;\;\dot{k}_j=-\sum_{p=5}^6A_{pj}h_j.
	\label{e:perturbeqn}
\end{equation}
For fixed planetary semimajor axes, the coefficients are constants, and the solution is given by a linear superposition of eigenmodes:
\begin{equation}
	\{h_j,k_j\}=\sum_pE_j^{(p)}\{\cos(g_pt+\beta_p),\sin(g_pt+\beta_p)\},
	\label{e:hkvector}
\end{equation}
where $g_p$ are the eigenfrequencies of the matrix $\mathbf{A}$ and $E_j^{(p)}$ are the corresponding eigenvectors; the amplitudes of the eigenvectors and the phases $\beta_p$ are determined by initial conditions.  
In our 2-planet model, the secular frequencies $g_5$ and $g_6$ depend on the masses of Jupiter, Saturn, and the Sun and on the semimajor axes of Jupiter and Saturn.

For the current semimajor axes and eccentricities of Jupiter and Saturn the Laplace-Lagrance theory gives frequency values $g_5=3.7\arcsec\yr^{-1}$ and $g_6=22.3\arcsec\yr^{-1}$, which are lower than the more accurate values given by \cite{Brouwer:1950p109} by $14\%$ and $20\%$, respectively~\citep{Laskar:1988p597}.
\cite{Brouwer:1950p109} achieved their more accurate solution by incorporating higher order terms in the disturbing function involving $2\lambda_5-5\lambda_6$, which arise due to Jupiter and Saturn's proximity to the 5:2 resonance (the so-called ``Great Inequality'').
By doing an accurate numerical analysis (described below), we found that the effect of the 5:2 resonance is only important over a very narrow range in Saturn's semimajor axis.  
More significant is the perturbation owing to the 2:1 near-resonance of Jupiter and Saturn.
\cite{Malhotra:1989p106} developed corrections to the Laplace-Lagrange theory to account for the perturbations from $n+1:n$ resonances in the context of the Uranian satellite system.  
Applying that approach to our problem, we find that the  2:1 near-resonance between Jupiter and Saturn leads to zeroth order corrections to the elements of the  $\mathbf{A}$ matrix\footnote{Corrections due to near-resonances are of order $e^{(n-1)}$, where $e$ is eccentricity and $n$ is the order of the resonance. 
The 5:2 is a third order resonance, so its effect is ${\cal O}(e^2)$.  
The 2:1 is a first order resonance, so that its effect does not depend on $e$. 
Therefore the discrepancy between linear theory and numerical analysis (or the higher order theory of \citeauthor{Brouwer:1950p109}) arising from the Great Inequality would be much less if Jupiter and Saturn were on more circular orbits, but the effect due to the 2:1 resonance would remain.}.  
Including these corrections, we determined the secular frequencies for a range of values of Saturn's semimajor axis; the result for $g_6$ is shown in Fig.~\ref{f:g6vsasat}a (dashed line).

We also calculated values for the eccentricity-pericenter eigenfrequencies by direct numerical integration of the full equations of motion for the two-planet, planar solar system.
In these integrations, Jupiter's initial semimajor axis was $5.2\AU$, Saturn's semimajor axis, $a_6$, was one of 233 values in the range $7.3$--$10.45\AU$, initial eccentricities of Jupiter and Saturn were $0.05$, and initial inclinations were zero.  
The initial longitude of pericenter and mean anomalies of Jupiter were $\varpi_{5,i}=15^\circ$ and $\lambda_{5,i}=92^\circ$, and Saturn were $\varpi_{6,i}=338^\circ$ and $\lambda_{6,i}=62.5^\circ$.
In each case, the planets' orbits were integrated for 100 myr, and a Fourier transform of the time series of the $\{h_j,k_j\}$ yields their spectrum of secular frequencies.
For regular (non-chaotic) orbits, the spectral frequencies are well defined and are readily identified with the frequencies of the secular solution.
The $g_6$ frequency as a function of Saturn's semimajor axis was obtained by this numerical analysis; the result is shown by the solid line in Fig.~\ref{f:g6vsasat}a.

The comparison between the numerical analysis and the Laplace-Lagrange secular theory indicates that the linear secular theory, including the corrections due to the 2:1 near-resonance, is an adequate model for the variation in $g_6$ as a function of $a_6$.  We adopted the latter for its convenience in the needed computations. 
The value of $a_{\nu_6}$ as a function of Saturn's semimajor axis was thus found by solving for the value of asteroid semimajor axis where $g_0=g_6$; $g_0$ was calculated using equation~(\ref{e:g0def}) and $g_6$ is the eigenfrequency associated with the $p=6$ eigenmode (at each value of Saturn's semimajor axis).  
The result is shown in Fig.~\ref{f:g6vsasat}b.

\begin{figure}[htb]
\centering
\plotone{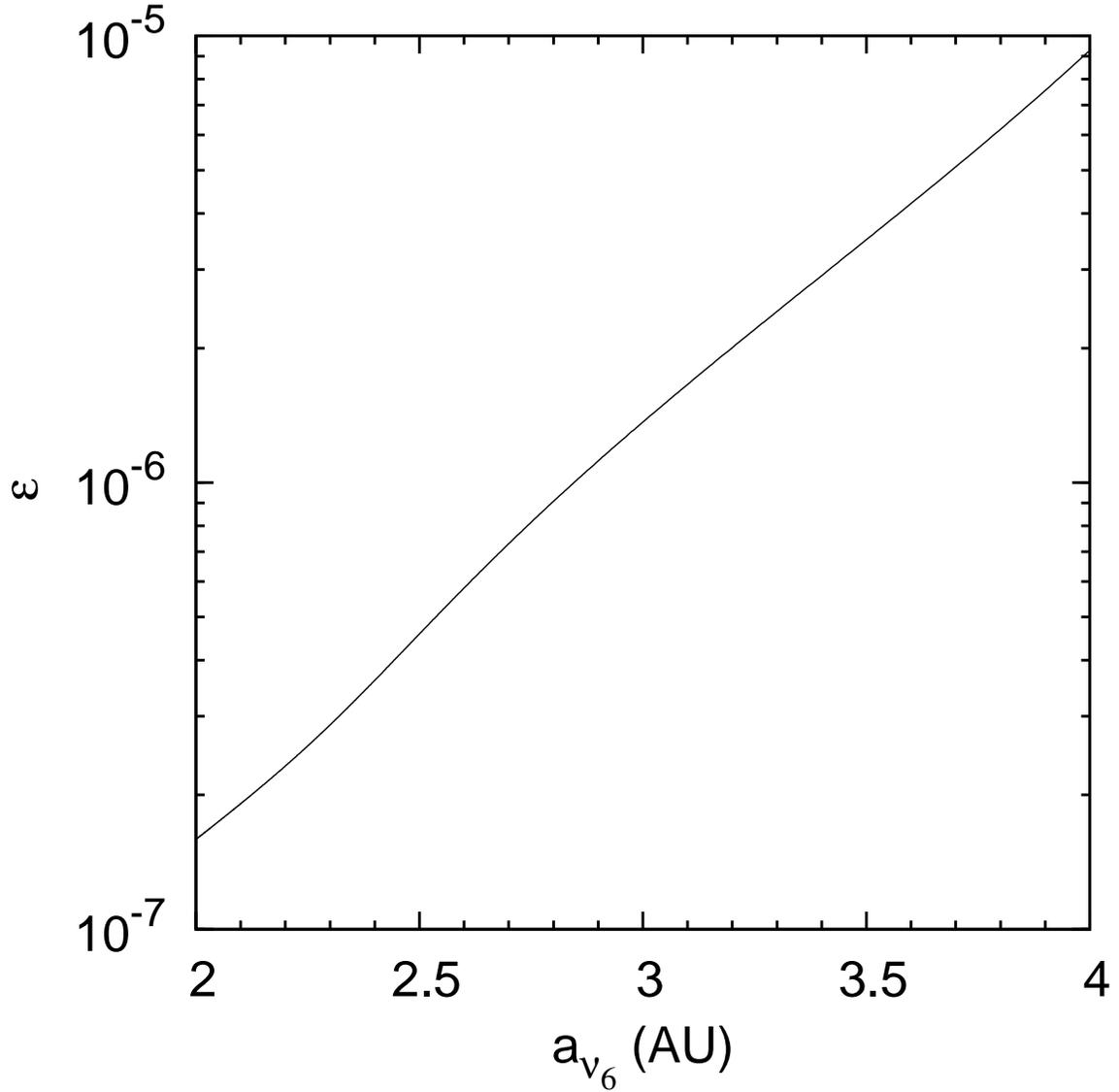}
\caption
{The value of the coefficient $\varepsilon$ defined by equation~(\ref{e:epsdef}) as a function of the zero inclination location of the $\nu_6$ resonance.
The values of $E_j^{(i)}$ were calculated using first order Laplace-Lagrange secular theory with corrections arising from the 2:1 Jupiter-Saturn mean motion resonance.
}
\label{f:epsvsanu6}
\end{figure}

We also used the analytical secular theory to calculate the eigenvector components $E_j^{(6)}$ in the secular solution of the 2-planet system, for each value of Saturn's semimajor axis.  
We adopted the same values for the initial conditions of Jupiter and Saturn as in the direct numerical integrations discussed above.  
Finally, we computed the values of the parameter $\varepsilon$ at each location $a_{\nu_6}$ of the secular resonance.  
The result is plotted in Fig.~\ref{f:epsvsanu6}.  
Despite the complexity of the computation, the result shown in Fig.~\ref{f:epsvsanu6} is approximated well by a simple exponential curve, 
$\varepsilon\approx3.5\times10^{-9}\exp(2a_{\nu_6}/\mathrm{AU})$, in the semimajor axis range $2<a_{\nu_6}/\mathrm{AU}<4$.

\subsection{Four test cases}
\label{sec:cases}
We checked the results of our analytical model against four full numerical simulations of the restricted four-body problem (the Sun-Jupiter-Saturn system with test particle asteroids) in which the test particles in the asteroid belt are subjected to the effects of a migrating Saturn.
The numerical integration was performed with an implementation of a symplectic mapping~\citep{Wisdom:1991p139,Saha:1992p42}, and the integration stepsize was $0.01\yr$.  
Jupiter and Saturn were the only massive planets that were integrated, and their mutual gravitational influence was included.  
The asteroids were approximated as massless test particles.  
The current solar system values of the eccentricity of Jupiter and Saturn were adopted and all inclinations (planets and test particles) were set to zero.
An external acceleration was applied to Saturn to cause it to migrate outward linearly and smoothly starting at $8.5\AU$ at the desired rate. 
As shown in Fig.~\ref{f:g6vsasat}b, the $\nu_6$ resonance was located at $3\AU$ at the beginning of the simulation, and swept inward past the current location of the inner asteroid belt in all simulations.
In each of the four simulations, 30 test particles were placed at $2.3\AU$ and given different initial longitudes of pericenter spaced $12^\circ$ apart.
The semimajor axis value of $2.3\AU$ was chosen because it is far away from the complications arising due to strong mean motion resonances.
The only parameters varied between each of the four simulations were the initial osculating eccentricities of the test particles, $e_i$, and the migration speed of Saturn, $\dot{a}_{6}$.  
The parameters explored were:
\renewcommand{\labelenumi}{\alph{enumi})}
\begin{enumerate}
\item $e_i=0.2$, $\dot{a}_{6}=1.0\AU\My^{-1}$;
\item $e_i=0.2$, $\dot{a}_{6}=0.5\AU\My^{-1}$;
\item $e_i=0.1$, $\dot{a}_{6}=1.0\AU\My^{-1}$;
\item $e_i=0.3$, $\dot{a}_{6}=1.0\AU\My^{-1}$.
\end{enumerate}
These values were chosen to illustrate the most relevant qualitative features.
The migration rates of $0.5\AU\My^{-1}$ and $1.0\AU\My^{-1}$ are slow enough so that the change in eccentricity is substantial, but not so slow that the non-linear effects at high eccentricity swamp the results.  
These test cases illustrate both how well the analytical model matches the numerical results, and where the analytical model breaks down. 

\begin{figure}[htb]
\plottwo{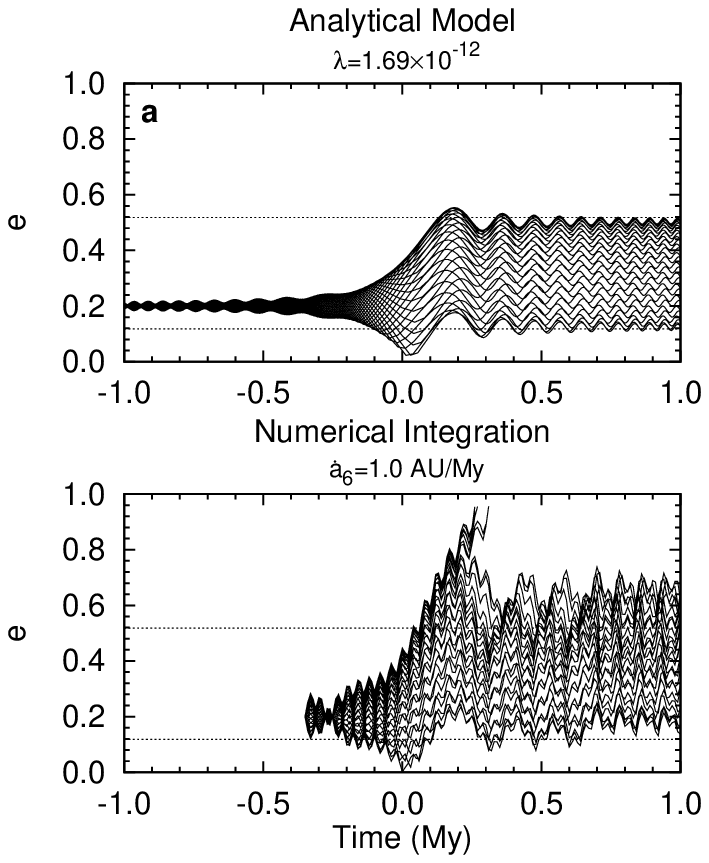}{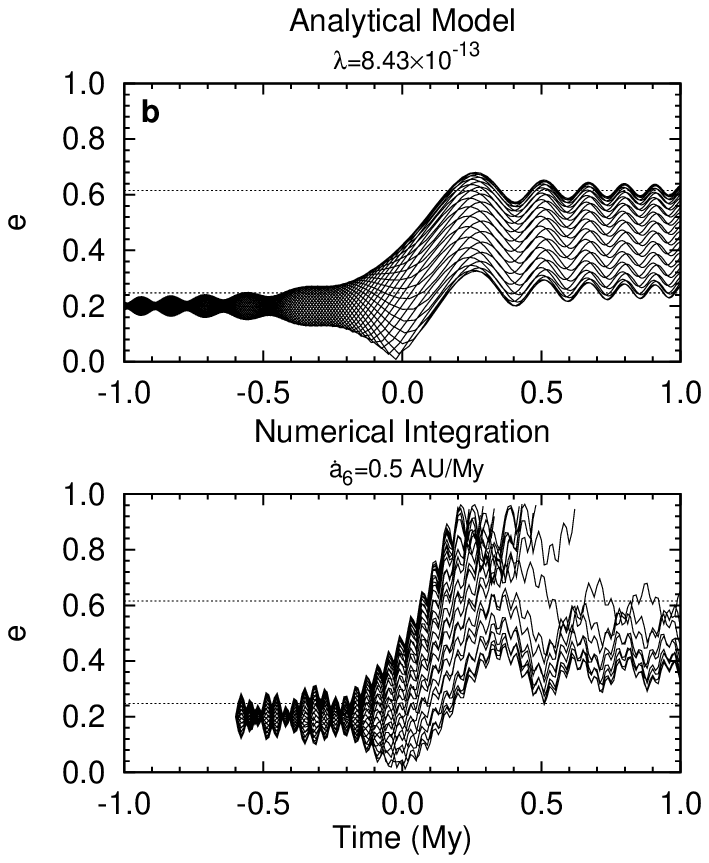}\\
\plottwo{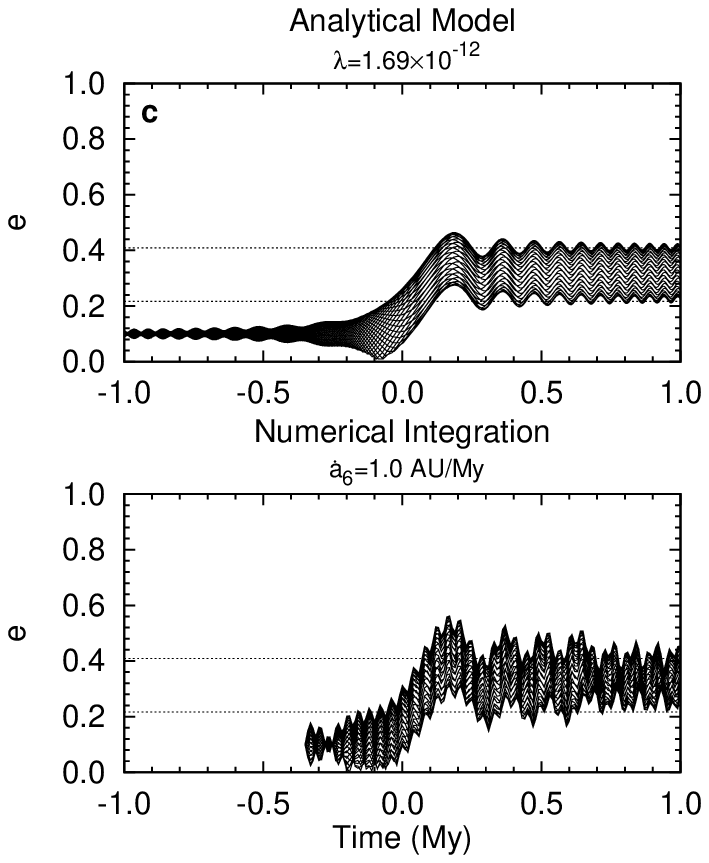}{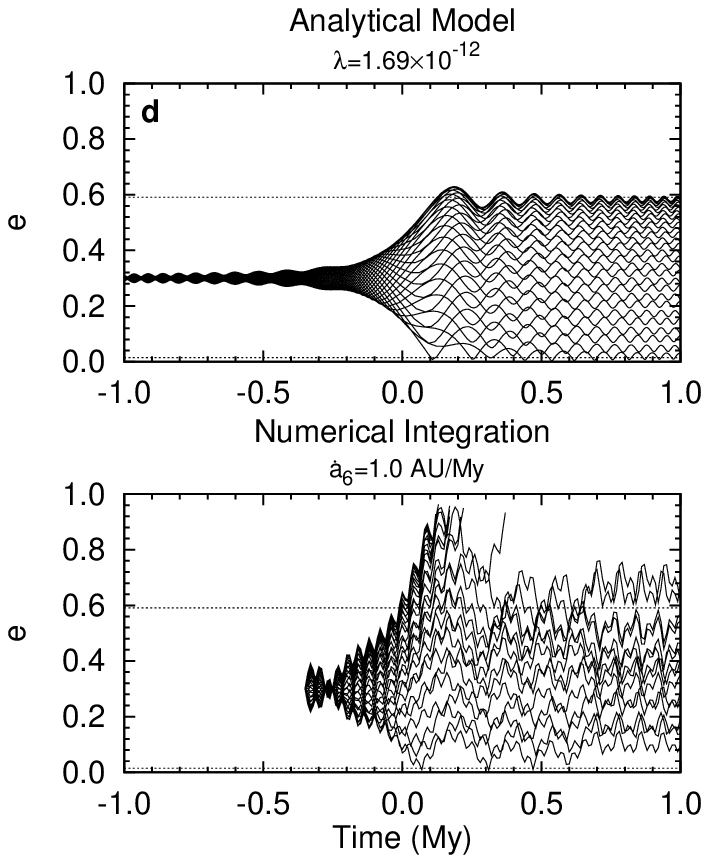}
\caption{Comparison between the numerical solution of the averaged equations (equations~\ref{e:dotxsweep}--\ref{e:dotysweep}) and full numerical integrations of test particles at $a=2.3\AU$.  
The dashed lines represent the envelope of the predicted final eccentricity, equation~(\ref{e:finalebounds}).
The values of $\lambda$ given are in the canonical unit system described in \S\ref{sec:sweeprate-toy_model}.
Each panel labeled {\bf a}--{\bf d} plots both an analytical theory result and a numerical integration result for each of the four test cases labeled a--d in section~\ref{sec:cases}.
The integrations were performed with Saturn starting at $8.5\AU$ and migrating outward linearly, while Jupiter remained fixed at $5.2\AU$. Jupiter and Saturn had their current eccentricities but zero inclination.
The thirty test particles in each numerical simulation were placed at $2.3\AU$ with zero inclination, but with longitudes of perihelion spaced $12^\circ$ apart.
Time zero is defined as the time when the $\nu_6$ resonance reached $2.3\AU$.
}

\label{f:phasetest}
\end{figure}

Two aspects of the analytical model were checked.
First, the perturbative equations of motion, equations~(\ref{e:dotxsweep}) and (\ref{e:dotysweep}), were numerically integrated, and their numerical solution compared with the numerical solution from the direct numerical  integration of the full equations of motion.  
For the perturbative solution, we adopted values for $\lambda$ that were approximately equivalent to the values of $\dot{a}_{6}$ in the full numerical integrations.
Second, the eccentricity bounds predicted by the analytical theory, equation~(\ref{e:finalebounds}), were compared with both numerical solutions.
The results of these comparisons for the four test cases are shown in Fig.~\ref{f:phasetest}.
We find that the analytically predicted values of the maximum and minimum final eccentricities (shown as horizontal dashed lines) are in excellent agreement with the final values of the eccentricities found in the numerical solution of the perturbative equations, and in fairly good agreement with those found in the full numerical solution. 
Not surprisingly, we find that the test particles in the full numerical integrations exhibit somewhat more complicated behavior than the perturbative approximation, and equation~(\ref{e:finalebounds}) somewhat underpredicts the maximum final eccentricity: this may be due to higher order terms in the disturbing function that have been neglected in the perturbative analysis and which become more important at high eccentricity; effects due to close encounters with Jupiter also become important at the high eccentricities.

\subsection{Comparison with observed asteroid eccentricities}
\label{sec:sweeprate-eccentricity_effect}

Does the eccentricity distribution of main belt asteroids retain features corresponding to the effects of the $\nu_6$ resonance sweeping? To answer this question, we need to know the main belt eccentricity distribution free of observational bias, and also relatively free of the effects of $\sim4\Gy$ of collisional evolution subsequent to the effects of planetary migration.
We therefore obtained the proper elements of the observationally complete sample of asteroids with absolute magnitude $H\leq10.8$ from the AstDys online data service~\citep{Knezevic:2003p74}; we excluded from this set the members of collisional families as identified by~\citet{Nesvorny:2006p2242}.  
These same criteria were adopted in~\cite{Minton:2010p3376} in a study of the long term dynamical evolution of large asteroids.
This sample of 931 main belt asteroids is a good approximation to a complete set of large asteroids that have been least perturbed by either dynamical evolution or collisional evolution since the epoch of the last major dynamical event that occurred in this region of the solar system; therefore this sample likely preserves best the post-migration orbital distribution of the asteroid belt.   
The proper eccentricity distribution of these asteroids is shown in Fig.~\ref{f:MBA-big-dist}.   
This distribution has usually been described in the literature by simply quoting its mean value (and sometimes a dispersion)~\citep{Murray:1999SSD,OBrien:2007p95}.
Our best fit single Gaussian distribution to this data has a mean, $\mu_e$ and standard deviation, $\sigma_e$, given by $\mu_e=0.135\pm0.00013$ and $\sigma_e=0.0716\pm0.00022$, and is plotted in Fig.~\ref{f:MBA-big-dist}.
However, we also note (by eye) a possible indication of a double-peak feature in the observed population. 
Our best fit double Gaussian distribution (modelled as two symmetrical Gaussians with the same standard deviation, but different mean values) to the same data has the following parameters:
\begin{align*}
	\mu'_{e,1}&=0.0846\pm0.00011,\\
	\mu'_{e,2}&=0.185\pm0.00012,\\
	\sigma_e'&=0.0411\pm0.00020.
\end{align*}
More details of how these fits were obtained are described in Appendix~\ref{sec:sweeprate-appendix-fitting}, where we also discuss goodness-of-fit of the single and double gaussian distributions. A Kolmogorov-Sinai test shows that there is only a 4.5\% probability that the observed eccentricities actually are consistent with the best-fit single Gaussian distribution, and a 73\% probability that they are consistent with the best-fit double Gaussian distribution; while the double gaussian is apparently a better fit to the data compared to the single gaussian, a K-S probability of only 73\% is quite far from a statistically significant level of confidence.  A dip test for multi-modality in the observed eccentricity distribution \citep{Hartigan:1985p3924} is also inconclusive;  the likely reason for this is that our data sample is not very large, and the separation between the two putative peaks is too small in relation to the dispersion. 
Nevertheless, we bravely proceed in the next section with the implications of a double-Gaussian eccentricity distribution, with the caveat that some of the conclusions we reach are based on this statistically marginal result.

\section{A constraint on Saturn's migration rate}
\label{sec:constraining_saturn_rate}
\begin{figure}[htb]
\plottwo{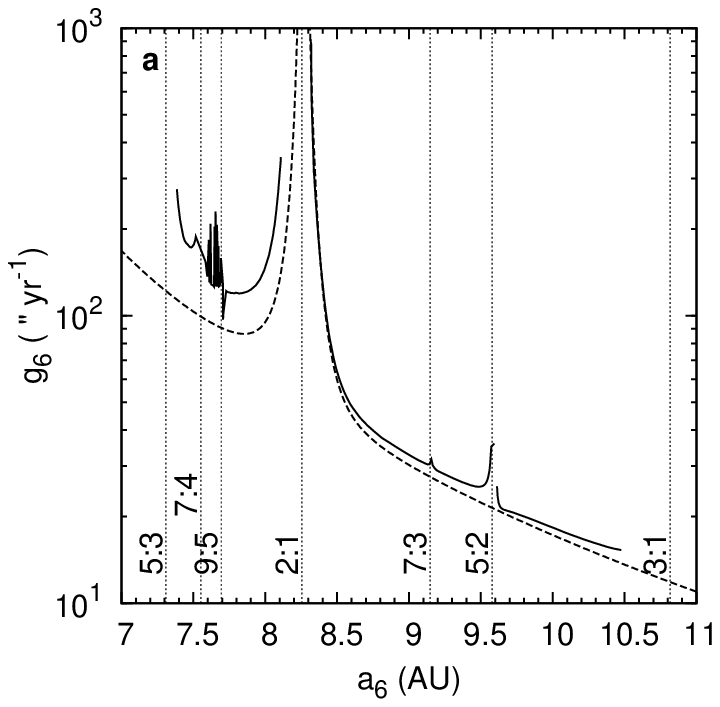}{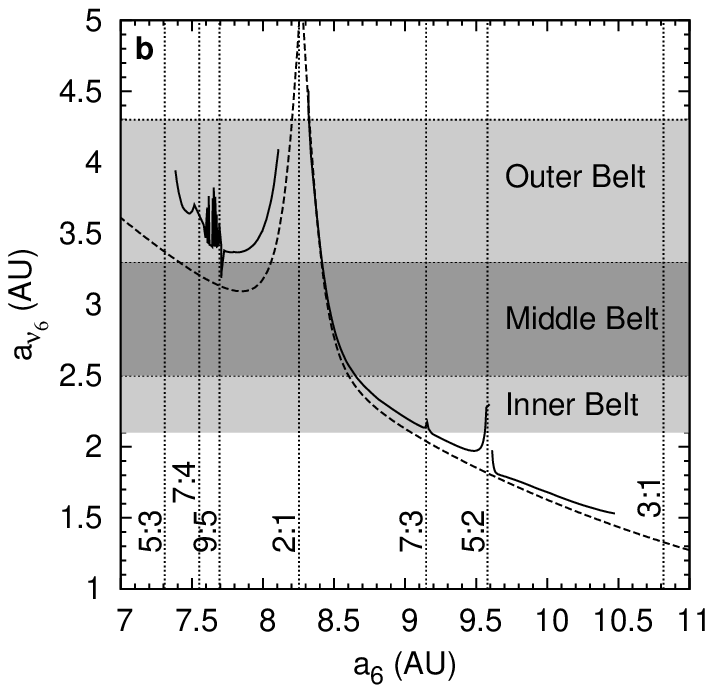}
\caption{
{\bf a)} The $g_6$ eigenfrequency as a function of Saturn's semimajor axis, for Jupiter fixed at $5.2\AU$.  
The dashed line shows the result from linear secular theory, with a correction for the effect of the near 2:1 mean motion resonance between Jupiter and Saturn~\citep{Malhotra:1989p106}.  The solid line shows the result from numerical spectral analysis of 233 solar system integrations (see text for explanation). The locations of Jupiter-Saturn MMRs which have an effect on the value of $g_6$ are indicated by vertical dotted lines. 
{\bf b)} The location of the $\nu_6$ resonance (at zero inclination) as a function of Saturn's orbit.  The frequencies $g_6$ and $g_0$ were calculated for each value of Saturn's semimajor axis, $a_6$, and then the location $a_{\nu_6}$ was determined by finding where $g_0-g_6=0$. 
The dashed line shows the result from linear secular theory, with a correction for the effect of the 2:1 near-MMR between Jupiter and Saturn~\citep{Malhotra:1989p106}. The solid line was obtained by using the $g_6$ eigenfrequencies obtained from spectral analysis of the 233 numerical integrations, as shown in ({\bf a}).}
\label{f:g6vsasat}
\end{figure}

By relating the $g_6$ secular frequency to the semimajor axis of Saturn, $\dot{g}_6$ can be related to the migration rate of Saturn, $\dot{a}_{6}$.
In this section only the effects of the sweeping $\nu_6$ secular resonance will be considered, and effects due to sweeping jovian mean motion resonances will be ignored.
Because Jupiter is thought to have migrated inward a much smaller distance than Saturn migrated outward during planetesimal-driven migration, the effects due to migrating jovian MMRs were likely confined to narrow regions near strong resonances~\citep{Minton:2009p280}.
In the inner asteroid belt between $2.1$--$2.8\AU$, these would include the 3:1 and 5:2 resonances, currently located at approximately $2.5\AU$ and $2.7\AU$, respectively.  
As shown in Fig.~\ref{f:g6vsasat}b, plausible parameters for the outward migration of Saturn would have allowed the $\nu_6$ to sweep across the entire inner asteroid belt.
Therefore the $\nu_6$ resonance would have been the major excitation mechanism across the $2.1$--$2.8\AU$ region of the main belt (and possibly across the entire main asteroid belt, depending on Saturn's pre-migration semimajor axis) during giant planet migration.

We used the results of our analytical model to set limits on the rate of migration of Saturn assuming a linear migration profile, with the caveat that many important effects are ignored, such as asteroid-Jupiter mean motion resonances, and Jupiter-Saturn mean motion resonances (with the exception of the 2:1 resonance). 
We have confined our analysis to only the region of the main belt spanning $2.1$--$2.8\AU$.  
Beyond $2.8\AU$ strong jovian mean motion resonance become more numerous.
Due to the high probability that the icy planetesimals driving planet migration would be ejected from the solar system by Jupiter, Jupiter likely migrated inward.
The migration of Jupiter would have caused strong jovian mean motion resonances to sweep the asteroid belt, causing additional depletion beyond that of the sweeping $\nu_6$ resonance~\citep{Minton:2009p280}.
A further complication is that sweeping jovian mean motion resonances may have also trapped icy planetesimals that entered the asteroid belt region from their source region beyond Neptune~\citep{Levison:2009p1637}.
The effects of these complications are reduced when we consider only the inner asteroid belt.
From Fig.~\ref{f:g6vsasat}b, we find that the $\nu_6$ would have swept the inner asteroid belt region between $2.1$--$2.8\AU$ when Saturn was between $\sim8.5$--$9.2\AU$.
Therefore the limits on $\dot{a}_6$ that we set using the inner asteroid belt as a constraint are only applicable for this particular portion of Saturn's migration history.

\begin{figure}[htb]
\includegraphics[scale=.75]{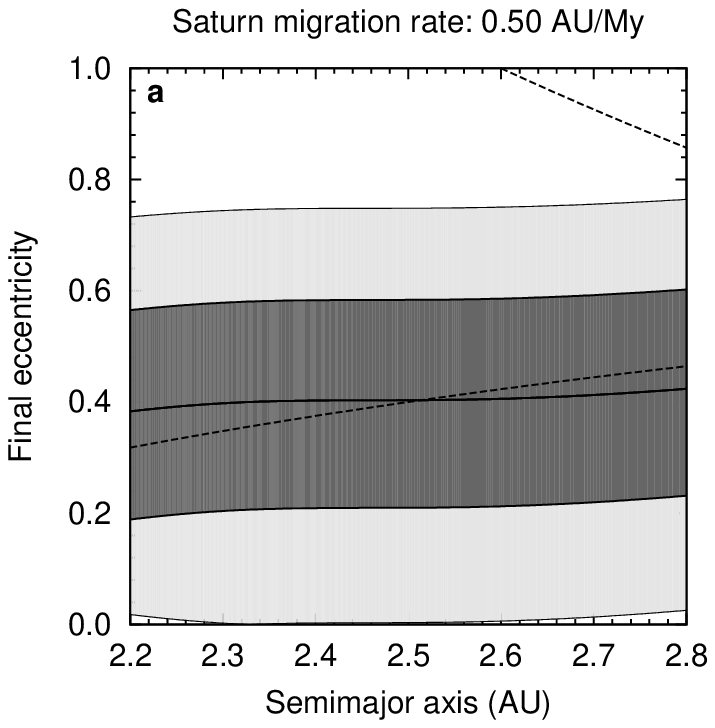}\includegraphics[scale=0.75]{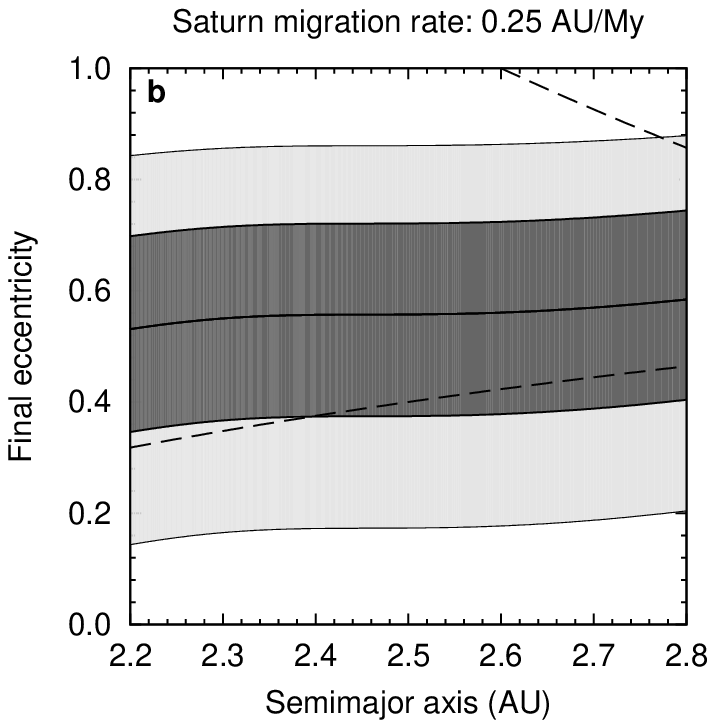}\includegraphics[scale=0.75]{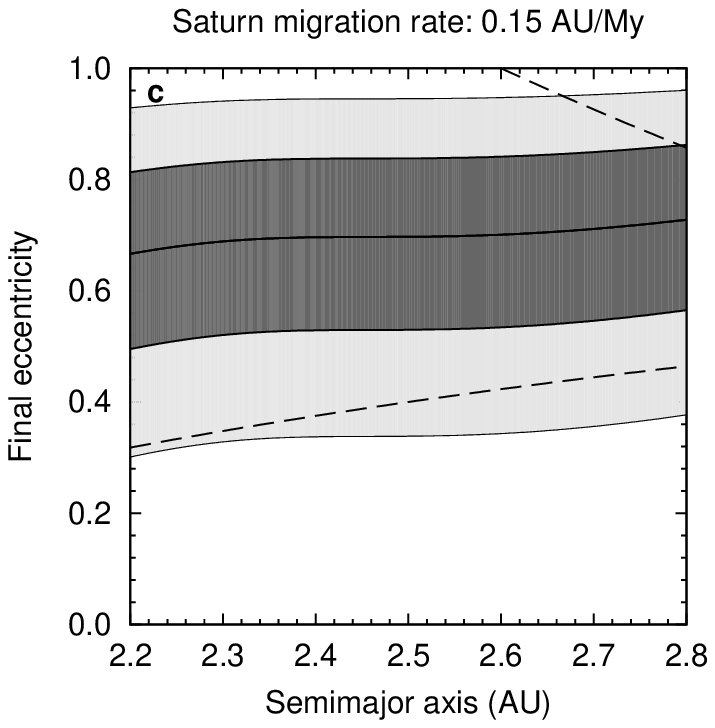}
\caption
{The final eccentricity of asteroids as a function of asteroid  semimajor axis and eccentricity for three different migration rates of Saturn, estimated from equation~(\ref{e:finalebounds}).  
Asteroids swept by the $\nu_6$ resonance can have a range of final eccentricities depending on their apsidal phase, $\varpi_i$.
The outermost shaded region demarcates the range of final eccentricities for asteroids with an initial eccentricity $e_i=0.4$.  
The innermost shaded region demarcates the range of final eccentricities for asteroids with an initial eccentricity $e_i=0.2$.
The solid line at the center of the shaded regions is the final eccentricity for an asteroid with an initial eccentricity $e_i=0$.  
}
\label{f:sweepratefigs}
\end{figure}

Our theoretically estimated final eccentricity as a function of initial asteroid semimajor axis and eccentricity is shown in Fig.~\ref{f:sweepratefigs} for three different adopted migration rates of Saturn.  
The larger the initial asteroid eccentricities, the wider the bounds in their final eccentricities.
If we adopt the reasonable criterion that an asteroid is lost from the main belt when it achieves a planet-crossing orbit (that is, crossing the orbits of either Jupiter or Mars) and that initial asteroid eccentricities were therefore confined to $\lesssim0.4$, then from Fig.~\ref{f:sweepratefigs} Saturn's migration rate must have been $\dot{a}_6\gtrsim0.15\AU\My^{-1}$ 
when the $\nu_6$ resonance was sweeping through the inner asteroid belt. 
Our results indicate that if Saturn's migration rate had been slower than $0.15\AU\My^{-1}$ when it was migrating across $\sim8.5$--$9.2\AU$, then the inner asteroid belt would have been completely swept clear of asteroids by the $\nu_6$ resonance.

%***

\begin{figure}[htb]
\centering
\plotone{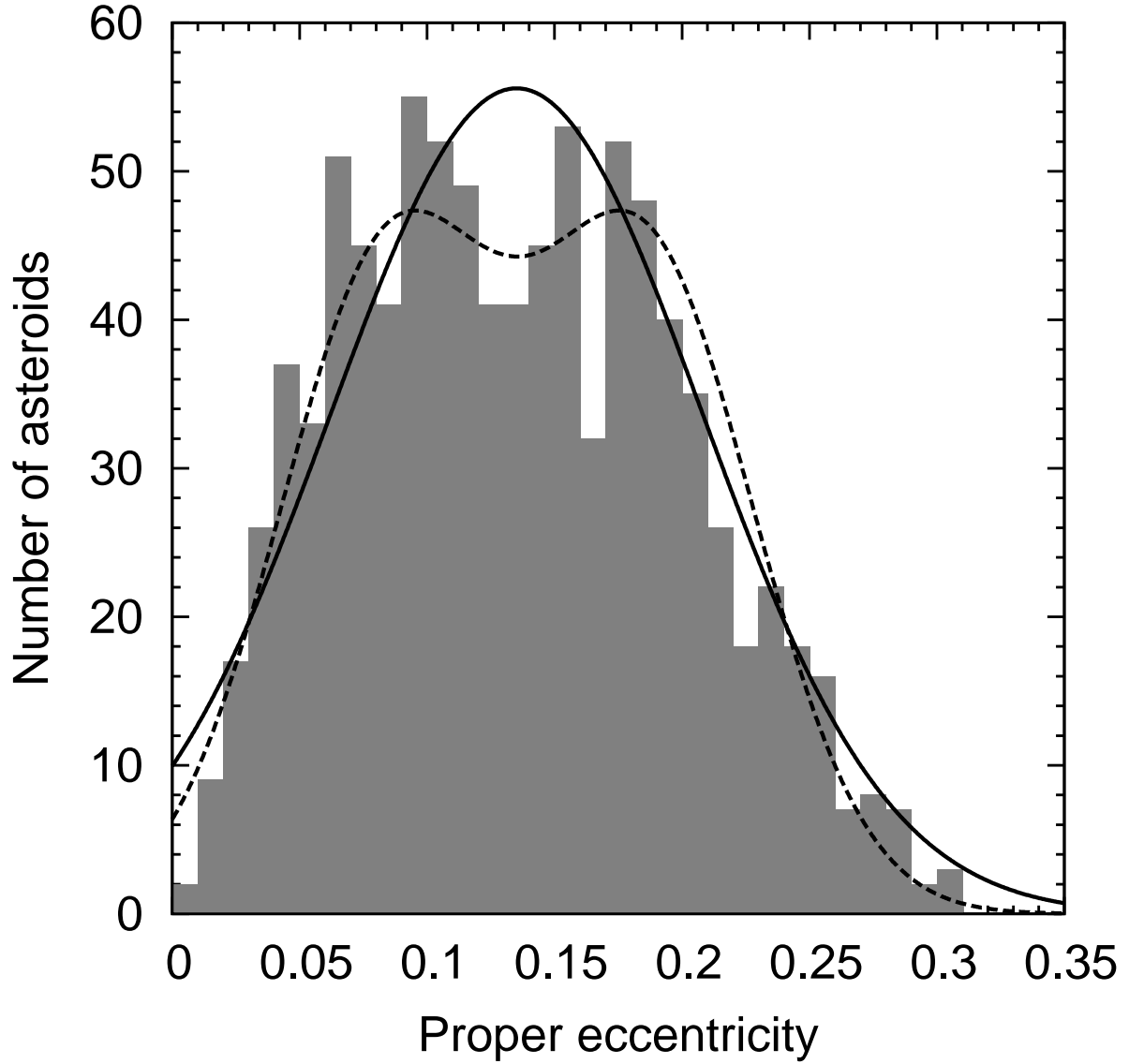} 
\caption
{
Proper eccentricity distribution of the 931 observed asteroids with absolute magnitude $H\leq10.8$, excluding members of collisional families.
The proper elements were taken from the AstDys online data service~\citep{Knezevic:2003p74}.
Family members identified by \cite{Nesvorny:2006p2242} were excluded.
The solid lines are the best fit Gaussian distribution to the observational data.
The dashed line is the best fit double-Gaussian distribution.
}
\label{f:MBA-big-dist}
\end{figure}

In light of our analysis and the observed dispersion of eccentricities in the asteroid belt (Fig.~\ref{f:MBA-big-dist}), we can also immediately conclude that the pre-migration asteroid belt between $2.1$ and $2.8\AU$ had significantly non-zero eccentricities.    
This is because, as discussed at the top of \S\ref{s:sweeping}, an initially cold asteroid belt swept by the $\nu_6$ resonance would either lose all its asteroids or none, and very low initial eccentricities would result in final asteroid eccentricities in a very narrow range of values (cf.~equation~\ref{e:finalebounds}), in contradiction with the fairly wide eccentricity dispersion that is observed.
This conclusion supports recent results from studies of planetesimal accretion and asteroid and planet formation that the asteroids were modestly excited at the end of their formation \citep[e.g.,~][]{Petit:2002p170}. 

In Appendix~\ref{sec:sweeprate-appendix-fitting} we show that the double-gaussian distribution is a slightly better fit to the main belt asteroid eccentricity distribution, 
but the statistical tests do not rule out a single-peaked distribution.  We boldly proceed with considering the implications of the double-peaked eccentricity distribution to further constrain the migration rate of Saturn, with the caveat that these results can only be said to be consistent with the observations, rather than uniquely constrained by them.

If the pre-sweeping asteroid belt had a Gaussian eccentricity distribution, then the lower peak of the post-sweeping asteroid belt should be equal to the lower bound of equation~(\ref{e:finalebounds}).
We use the analytical theory to make a rough estimate of the parameter $\lambda$ (and hence $\dot a_6$) that would yield a final distribution with lower peak near $0.09$ 
and upper peak near $0.19$ (which is similar to the best-fit double Gaussian in Fig.~\ref{f:MBA-big-dist}). 
 
Applying equation~(\ref{e:finalebounds}), we see that there are two possible solutions: 
$\langle e_i\rangle=0.14$,$\delta_e=0.05$ and 
$\langle e_i\rangle=0.05$,$\delta_e=0.14$.  
A corresponding migration rate of Saturn can be estimated from the value of $\delta_e$ using equation~(\ref{e:deltaedef}), and the parameter relationships plotted in 
Figs.~\ref{f:g6vsasat} and \ref{f:epsvsanu6}.
The former solution ($\delta_e=0.05$) requires a migration rate for Saturn of $\dot{a}_6=30\AU\My^{-1}$.
We mention this implausible solution here for completeness, but we will not discuss it any further.
The latter solution ($\delta_e=0.14$) requires a migration rate for Saturn of $\dot{a}_6=4\AU\My^{-1}$. 
We dub this solution the ``cold belt'' solution.
This rate is comparable to the rates of planet migration found in the ``Jumping Jupiter'' scenario proposed by \cite{Brasser:2009p3022}.
A third solution exists if we consider that eccentricities in the main belt are restricted by the orbits of Mars and Jupiter on either side, such that stable asteroid orbits do not cross the planetary orbits.  
This limits asteroid eccentricities to values such that neither the aphelion of the asteroid crosses the perihelion distance of Jupiter, nor the perihelion of the asteroid crosses the aphelion distance of Mars.
Maximum asteroid eccentricity is therefore a function of semimajor axis, where $e_{max}=\min(1-Q_{Mars}/a-1,q_{Jupiter}/a-1)$, where $Q$ and $q$ are planet aphelion and perihelion, respectively, and $a$ is the semimajor axis of the asteroid.
In this case, an initial single Gaussian eccentricity distribution with a mean greater than $\sim0.3$ would be severely truncated,  
therefore we need only fit the lower peak of the double Gaussian distribution at $e=0.09$.  
Applying equation~(\ref{e:finalebounds}), we find that $\delta_e=0.21$ provides a good fit. 
The corresponding migration rate of Saturn is $\dot{a}_6=0.8\AU\My^{-1}$.
We dub this solution the ``hot belt'' solution.

\begin{figure}[htb]
\plottwo{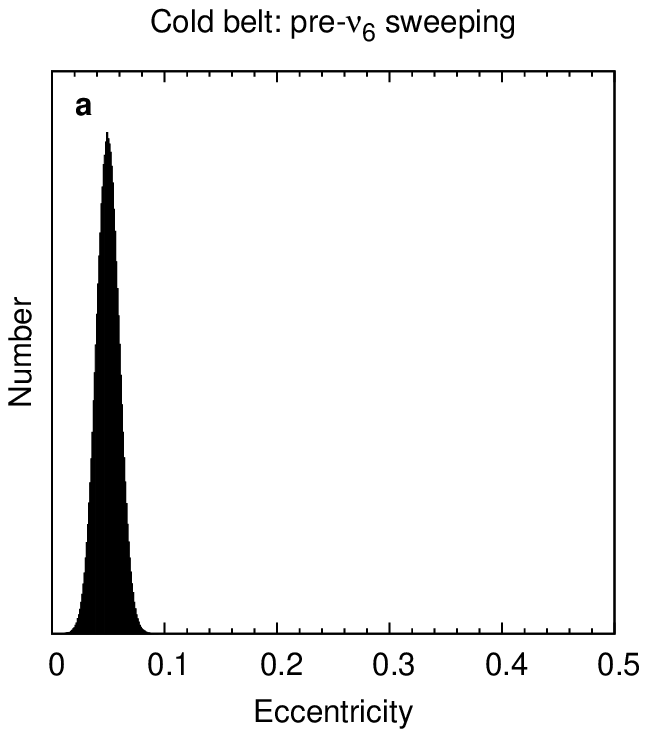}{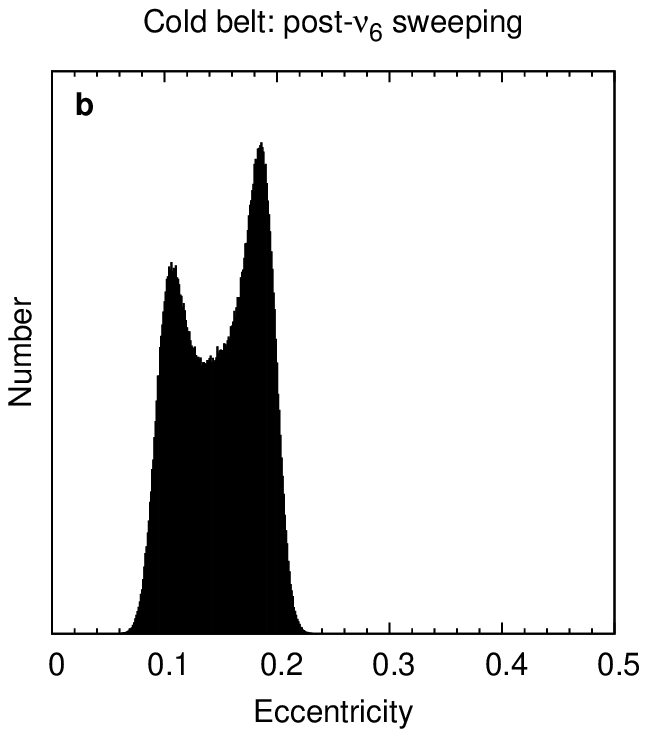}\\
\plottwo{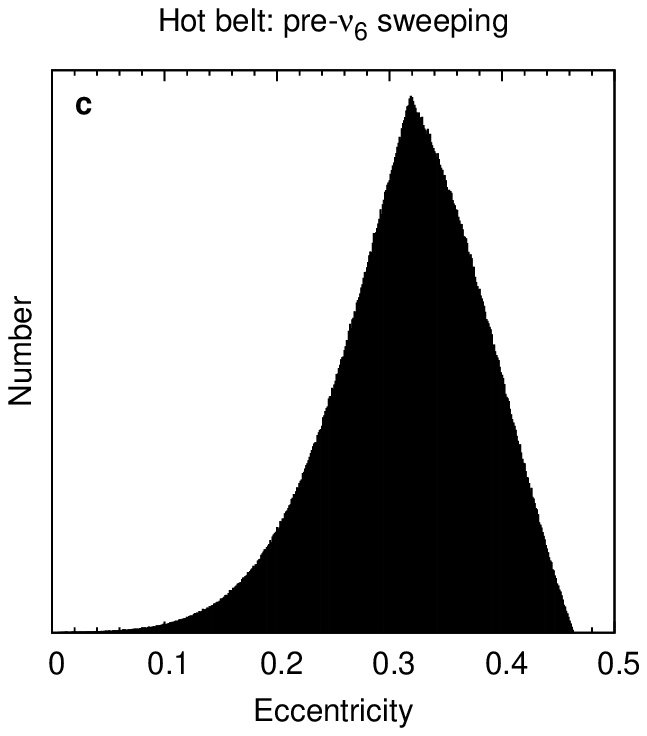}{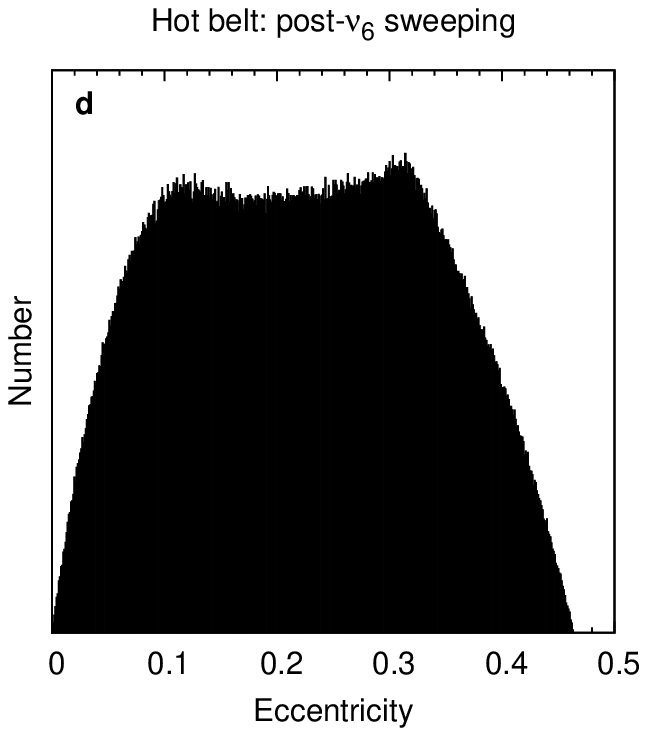}
\caption
{The effects of the sweeping $\nu_6$ resonance on an ensemble of fictitious asteroids with semimajor axes $2.1$--$2.8\AU$ and a uniform distribution of pericenter longitudes.
({\bf a}) Initial distribution of eccentricities, with mean $0.05$ and standard deviation $0.01$ (the ``cold belt'' solution).
({\bf b}) The final distribution of eccentricities after $\nu_6$ sweeping, estimated with the analytical theory (equation~(\ref{e:Jf})), for $\dot{a}_6=4.0\AU\My^{-1}$.
The two peaks in the final eccentricity distribution are at approximately the same values as the observed peaks in the main asteroid belt eccentricity distribution shown in Fig.~\ref{f:MBA-big-dist}.
({\bf c}) Initial distribution of eccentricities, with mean $0.40$ and standard deviation $0.1$ (the ``hot belt'' solution).  Asteroids with eccentricities greater than the Mars-crossing value were discarded.
({\bf d}) The final distribution of eccentricities after $\nu_6$ sweeping for $\dot{a}_6=0.8\AU\My^{-1}$.
The final distribution in {\bf d} is depleted by a factor of $2.3$ from the initial distribution in {\bf c}.
The ordinates in the four panels are not to the same scale.
}
\label{f:tpdist}
\end{figure}

We illustrate the two possible solutions for an ensemble of hypothetical asteroids having semimajor axes uniformly distributed randomly in the range $2.1\AU$ to $2.8\AU$.   
In Fig.~\ref{f:tpdist}a the initial eccentricity distribution is modeled as a Gaussian distribution with a mean $\langle e_i\rangle=0.05$ and a standard deviation of $0.01$.
This initial standard deviation was chosen so that the final standard deviation would be the same as that of the observed main belt. 
Fig.~\ref{f:tpdist}b shows the eccentricity distribution after $\nu_6$ resonance sweeping has occurred due to the migration of Saturn at a rate of $4\AU\My^{-1}$.
The final distribution was calculated with equation~(\ref{e:Jf}); we used values of $\varepsilon$ shown in Fig.~\ref{f:epsvsanu6}, and the value of $\lambda$ was calculated with the aid of Fig.~\ref{f:g6vsasat} which relates the value of $g_p$ to the semimajor axis of Saturn. 
As expected, when an ensemble of asteroids with a single-peaked eccentricity distribution is subjected to the sweeping secular resonance, the result is a double-peaked eccentricity distribution.   
Because of the slight bias towards the upper limit of the eccentricity excitation band, proportionally more asteroids are found in the upper peak.

In Fig.~\ref{f:tpdist}c the initial eccentricity distribution is modeled as a truncated Gaussian: a Gaussian with mean $\langle e_i\rangle=0.4$ and standard deviation $0.1$, but truncated at the semimajor axis--dependent Mars-crossing value.
We used equation~(\ref{e:Jf}) to calculate the eccentricity distribution of this hypothetical ensemble after $\nu_6$ resonance sweeping with $\dot a_6=0.8\AU\My^{-1}$.  
Again, allowing that only those asteroids whose final eccentricities are below the Mars-crossing value will remain, the resulting post-migration eccentricity distribution is shown in Fig.~\ref{f:tpdist}d.
In this case, we find the lower peak at the same eccentricity value as the lower peak in the observed main belt distribution (see Fig.~\ref{f:MBA-big-dist}).  

In both cases of possible solutions (initially cold main belt with $\langle e_i\rangle=0.05$ and $\dot{a}_6=4\AU\My^{-1}$; and initially hot main belt with $\langle e_i\rangle=0.4$ and $\dot{a}_6=0.8\AU\My^{-1}$), the theoretical models yield an excess of asteroids with eccentricities greater than $0.2$ than in the observed main belt. 
However, as shown by \cite{Minton:2010p3376}, on gigayear timescales, the $e\gtrsim0.2$ population of the asteroid belt is dynamically more unstable than the $e\lesssim0.2$ population.
Thus, both solutions may be consistent with the observations, as post-sweeping dynamical erosion could result in a final eccentricity distribution resembling more closely the observed distribution.

The estimates of Saturn's migration rate quoted above depend strongly on the eccentricities of the giant planets during their migration.  
In deriving the above estimates, we adopted the present values of the giant planets' orbital eccentricities.
The $\nu_6$ resonance strength coefficient $\varepsilon$ (equation~\ref{e:epsdef}) is proportional to the amplitude of the $p=6$ mode, which is related to the eccentricities of the giant planets (namely Saturn and Jupiter).
From equation~(\ref{e:hkvector}), and the definition $e_j(\sin\varpi_j,\cos\varpi_j)\equiv(h_j,k_j)$, the value of $E_j^{(p)}$ is a linear combination of the eccentricities of the giant planets.
Because Saturn is the planet with the largest amplitude of the $p=6$ mode, from equation~(\ref{e:Jf}) the relationship between the sweep rate and the value of Saturn's eccentricity is approximately $\lambda_{min}\propto e_6^2$.
Therefore, to increase the limiting timescale by a factor of ten would only require that the giant planets' eccentricities were $\sim0.3\times$ their current value (i.e., $e_{5,6}\approx0.015$).

\section{Conclusion and Discussion}
\label{sec:sweeprate-conclusion}
Based on the existence of the inner asteroid belt, we conclude that Saturn's migration rate must have been $\gtrsim0.15\AU\My^{-1}$ as Saturn migrated from $8.5$ to $9.2\AU$ (as the $\nu_6$ resonance migrated from 2.8 AU to 2.1 AU).   
Migration rates lower than $\sim0.15\AU\My^{-1}$ would be inconsistent with the survival of any asteroids in the inner main belt, as the $\nu_6$ secular resonance would have excited asteroid eccentricities to planet-crossing values.
This lower limit for the migration rate of Saturn assumes that Jupiter and Saturn had their current orbital eccentricities; the migration rate limit is inversely proportional to the square of the amplitude of the $g_6$ secular mode; if Jupiter and Saturn's eccentricities were $\sim0.3\times$ their current value (i.e., $e_{5,6}\approx0.015$) during the planet migration epoch, the limit on the migration rate decreases by a factor of $\sim10$.  (This caveat also applies to the migration rate limits quoted below.)

Our analysis of secular resonance sweeping predicts that a single-peaked eccentricity distribution will be transformed into a double-peaked eccentricity by secular resonance sweeping.   We find that the observed eccentricity distribution of asteroids may be consistent with a double-peaked distribution function, although we acknowledge that the statistics are poor.  We used a  double-Gaussian function to model the observed asteroid eccentricity distribution to set even tighter, albeit model-dependent, constraints on the migration rate of Saturn.  

We identified two possible migration rates that depend on the pre-migration dynamical state of the main asteroid belt.  
The first, the ``cold primordial belt'' solution, has an asteroid belt with an initial eccentricity distribution modeled as a Gaussian with $\langle e_i\rangle=0.05$; Saturn's migration rate of $4\AU\My^{-1}$ yields a final eccentricity distribution consistent with the observed asteroid belt.
The second, the ``hot primordial belt'' solution, has an asteroid belt with an initial eccentricity distribution modeled as a Gaussian with $\langle e_i\rangle=0.4, \sigma_e=0.1$, but truncated above the Mars-crossing value of eccentricity; in this case, Saturn's migration rate of $0.8\AU\My^{-1}$ is generally consistent with the observed asteroid belt.

Each of these solutions has very different implications for the primordial excitation and depletion of the main asteroid belt.
The cold belt solution, with $\dot{a}_6=4\AU\My^{-1}$, would lead to little depletion of the asteroid belt during giant planet migration, as the $\nu_6$ resonance would be unable to raise eccentricities to Mars-crossing values.  
This implies an initial dynamically quite cold asteroid belt with not more than $\sim2$ times the mass of the current main belt; the latter estimate comes from accounting for  dynamical erosion over the age of the solar system~\citep{Minton:2010p3376}.  

The hot belt solution, with $\dot{a}_6=0.8\AU\My^{-1}$, would lead to loss of asteroids directly due to excitation of eccentricities above planet-crossing values, by about a factor of $\sim2$.
In this case, the main asteroid belt was more dynamically excited prior to resonance sweeping than we find it today.
This implies much greater loss of asteroids prior to $\nu_6$-sweeping, as the peak of the eccentricity distribution would be near the Mars-crossing value, and subject to strong dynamical erosion~\citep{Minton:2010p3376}.

Each of these solutions has different implications for the model that the Late Heavy Bombardment of the inner solar system is linked to the epoch of planetesimal-driven giant planet migration~\citep{Gomes:2005p51,Strom:2005p80}.  These implications will be explored in a future paper.

We remind the reader that in order to elucidate the effects of $\nu_6$ resonance sweeping, we have made a number of simplifying assumptions to arrive at an analytically tractable model. 
These simplifications include neglecting the effects of planets other than Jupiter and Saturn, the effects of sweeping jovian mean motion resonances on asteroids, the effects of a presumed massive Kuiper belt during the epoch of planet migration, and the self-gravity and collisional interactions of a previously more massive asteroid belt. 
In addition, our analysis was carried out in the planar approximation, thereby neglecting any eccentricity-inclination coupling effects. 
These neglected effects can be expected to reduce somewhat the lower limit on Saturn's migration speed that we have derived, because in general they would reduce the effectiveness of the $\nu_6$ in exciting asteroid eccentricities.
Perhaps more importantly, giant planet migration would also lead to the sweeping of the main asteroid belt by the $\nu_{16}$ inclination secular resonance~\citep{Williams:1981p532} whose effects could be used to infer additional constraints.
% ***

%***
Recently, \cite{Morbidelli:2010p5466} found through numerical simulations with slow rates of planet migration (e-folding timescales exceeding 5 Myr) that the surviving asteroids in the main belt tend to be clumped around mean motion resonances. 
The semimajor axis distribution of survivors is found very different from the observed distribution for asteroids, and also that a large proportion of survivors have inclinations above 20$^\circ$, both inconsistent with the current main asteroid belt; comparisons made with the unbiased main belt asteroid distributions as described in~\cite{Minton:2009p280}. 
This is likely due to the effects of the sweeping $\nu_{16}$ inclination-longitude of ascending node secular resonance, analogous to the sweeping $\nu_6$ eccentricity-pericenter secular resonance that we analyzed in the present study.
While the effects of the sweeping $\nu_{16}$ resonance are analogous to the $\nu_6$, only affecting inclinations instead of eccentricities, a full analysis of the asteroid belt inclinations is beyond the scope of the present work, but will be explored in a future study.

A number of other studies have derived limits on the speed of planetesimal-driven giant planet migration.
\cite{MurrayClay:2005p209} exclude an $e-$folding migration timescale $\tau\leq 1\My$ to $99.65\%$ confidence based on the lack of a large observed asymmetry in the population of Kuiper belt objects in the two libration centers of the 2:1 Neptune mean motion resonance.
\cite{Boue:2009p1626} exclude $\tau\leq 7\My$ based on the observed obliquity of Saturn.  
The latter lower limit on the migration timescale is slightly incompatible with the lower limit on the rate of Saturn's migration of $\dot{a}_6>0.15\AU\My^{-1}$ we derive based on the existence of the inner asteroid belt.  
One way these can be reconciled is if Saturn's orbital eccentricity were a factor $\sim2$ smaller than its present value as it migrated from 8.5~AU to 9.2~AU; then, some mechanism would need to have increased Saturn's eccentricity up to its present value by the time Saturn reached its present semimajor axis of $\sim9.6\AU$.  

%***

\acknowledgements
The authors would like to thank the anonymous reviewer and the editor Eric Feigelson for useful comments.
This research was supported in part by NSF grant no.~AST-0806828 and NASA:NESSF grant no.~NNX08AW25H.
The work of David Minton was additionally partially supported by NASA NLSI/CLOE research grant no.~NNA09DB32A

\appendix
\section{The Main belt eccentricity distribution}
\label{sec:sweeprate-appendix-fitting}
The binned eccentricity distribution may be modeled as a Gaussian probability distribution function, given by:
\begin{equation}
	p(x)=\frac{1}{\sigma\sqrt{2\pi}}\exp\left[-\frac{(x-\mu)^2}{2\sigma^2}\right],
	\label{e:gaussian}
\end{equation}
where $\sigma$ is the standard deviation, $\mu$ is the mean, and $x$ is the random variable; in our case $x$ is the eccentricity.
With an appropriate scaling factor, equation~(\ref{e:gaussian}) can be used to model the number of asteroids per eccentricity bin. 
However, rather than fit the binned distribution, we instead perform a least squares fit of the unbinned sample to the Gaussian cumulative distribution function given by:
\begin{equation}
	P(x)=\frac{1}{2}+\frac{1}{2}\erf\left(-\frac{x-\mu}{\sigma\sqrt{2}}\right).
	\label{e:cdf}
\end{equation}
For the eccentricities of our sample of 931 main belt asteroids, the best fit parameters are:
\begin{align*}
	\mu_e&=0.135\pm0.00013,\\
	\sigma_e&=0.0716\pm0.00022.
\end{align*}

We also fit the data to a double-Gaussian distribution,
\begin{equation}
	p_2(x)=\frac{A'}{\sigma'\sqrt{2\pi}}\left\{\exp\left[-\frac{(x-\mu'_1)^2}{2\sigma'^2}\right]+\exp\left[-\frac{(x-\mu'_2)^2}{2\sigma'^2}\right]\right\}.
	\label{e:doublegaussian}
\end{equation}
The cumulative distribution function for equation~(\ref{e:doublegaussian}) is 
\begin{equation}
	P_2(x)=\frac{1}{2}+\frac{1}{4}\left[\erf\left(-\frac{x-\mu'_1}{\sigma\sqrt{2}}\right)+\erf\left(-\frac{x-\mu'_2}{\sigma\sqrt{2}}\right)\right].
	\label{e:doublecdf}
\end{equation}
For the eccentricities of our sample of 931 main belt asteroids, we performed a least squares fit to equation~(\ref{e:doublecdf}) and obtained the following best-fit parameters:  
\begin{align*}
	\mu'_{e,1}&=0.0846\pm0.00011,\\
	\mu'_{e,2}&=0.185\pm0.00012,\\
	\sigma_e'&=0.0411\pm0.00020.
\end{align*}

We evaluated the goodness of fit using the Kolmogorov-Smirnov (K-S) test.
The K-S test determines the probability that two distributions are the same, or in our case how well our model distributions fit the observed data~\citep{Press:1992p1610}.
The K-S test compares the cumulative distribution of the data against the model cumulative distribution function. 
We found that our asteroid sample has a probability of $4.5\times10^{-2}$ that it comes from the best fit single Gaussian (equation~(\ref{e:cdf})), but a probability of $0.73$ that it comes from the double-Gaussian (equation~(\ref{e:doublecdf})).  Therefore, the K-S tests indicate that the double-Gaussian is a better fit to the data than the single-Gaussian.

We performed Hartigan's dip test~\citep{Hartigan:1985p3924} to test whether the observational data is consistent with a multi-peaked distribution.
Hartigan's dip test calculates the probability that the distribution being tested has a single peak.
Applying Hartigan's dip test to a given distribution yields in a test statistic; together with the sample size, the test statistic is matched to a p-value range in a precomputed table provided by~\cite{Hartigan:1985p3924}.
The p-value is a measure of the probability that the distribution actually has only one peak (the null-hypothesis, for this problem).
The smaller the calculated p-value, the less likely is the distribution to have a single peak and the more likely it is to have at least two significant peaks.
A p-value of $<0.05$ indicates that the null-hypothesis is very unlikely, and that the given distribution has more than one peak.
We applied the dip test to the eccentricity distribution of our sample of 931 main belt asteroids; we determined that the test statistic is $0.0107$.   
From \cite{Hartigan:1985p3924}, this corresponds to a p-value range of $0.6<p<0.7$ (based on a sample size of 1000). 
This indicates that the Hartigan dip test does not rule out the null hypothesis, i.e., a single-peaked distribution cannot be ruled out.

To further aid the interpretation of this test, we compare this result to that obtained by applying Hartigan's dip test to synthetic distributions that were explicitly double-peaked by construction.  
Ten model distributions (of 1000 eccentricity values) were generated from the double-gaussian function of equation~(\ref{e:doublecdf}), with parameter values $\mu_1=0.0846$, $\mu_2=0.185$, and $\sigma=0.0411$ (same parameter values as the best-fit for our sample of asteroids).
Only the random seed was varied between each model distribution.
Applying the Hartigan dip test, we find that the test statistic ranged between $0.00852$ and $0.0126$, corresponding to p-values between $0.98$ and $0.3$.  
This means that, according to the dip test, the null hypothesis -- i.e., a single-peaked distribution -- could not be ruled out for any of the model distributions (since none had $p<0.05$), despite the fact that they were each drawn from an explicitly double-peaked distribution by construction.
This illustrates that even with a sample size of nearly 1000, the dispersion in eccentricity around the two peaks is too large in comparison to the distance between the peaks, that the dip test is not sufficiently sensitive to detect the double-peaked underlying distribution.
We interpret this to mean that the results of both the K-S test and Hartigan's dip test indicate that the main asteroid belt eccentricity distribution is consistent with being drawn from a double-peaked distribution, but that this cannot be definitely shown.

\bibliographystyle{apj} 
%\bibliography{allrefs}

\clearpage

\begin{figure}[htb]
\centering
\plotone{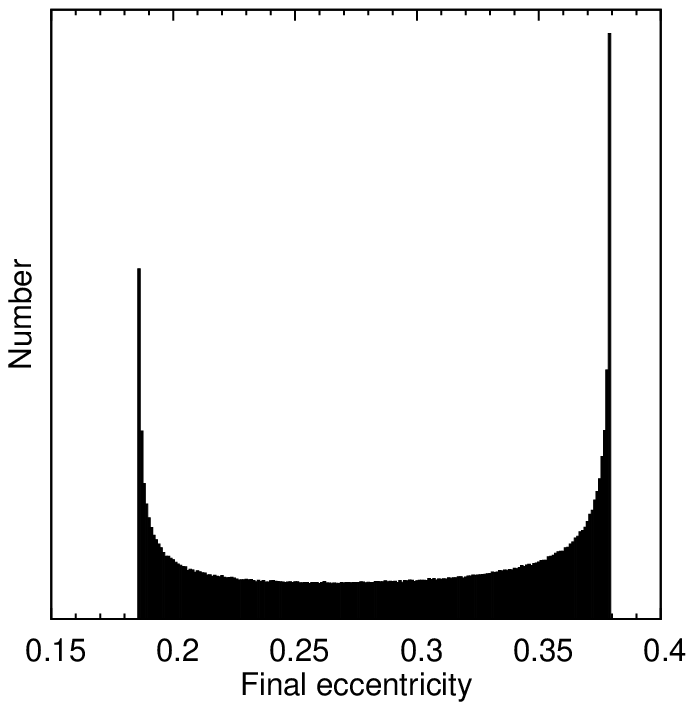}
\caption
{The final eccentricity distribution of an ensemble of particles, all having initial eccentricity $e_i=0.1$ but uniformly distributed values of the phase angle $\varpi_i$.
The effect due to the sweeping $\nu_6$ resonance was modeled using equation~(\ref{e:Jf}), with parameters chosen to simulate asteroids at $a=2.3\AU$, and with $\dot{a}_6=1\AU\My^{-1}$.
}
\label{f:tpideal}
\end{figure}

\end{document}